\documentclass[copyright,creativecommons]{eptcs}
\providecommand{\event}{QAPL 2013} 
\usepackage{amssymb}
\usepackage{amssymb}
\usepackage{amsmath}
\usepackage{amsfonts}
\usepackage{multirow}
\usepackage{xspace}
\usepackage{paralist}
\usepackage{algorithmic}
\usepackage{pseudocode}
\usepackage{stmaryrd}
\usepackage{color}
\usepackage{graphicx}
\usepackage{pgf}
\usepackage{tikz}
\usepackage{subfig}
\usetikzlibrary{arrows,positioning,automata,shadows,fit,shapes}
\usepackage[latin1]{inputenc}
\usepackage{verbatim}
\newtheorem{definition}{Definition}

\newtheorem{example}{Example}

\title{Quantitative Security Analysis for Multi-threaded Programs}
\author{Tri Minh Ngo
\institute{University of Twente, Netherlands}
\email{triminhngo@gmail.com}
\and
Marieke Huisman
\institute{University of Twente, Netherlands}
\email{Marieke.Huisman@ewi.utwente.nl}}

\def\titlerunning{Quantitative Security Analysis for Multi-threaded Programs}
\def\authorrunning{T.M. Ngo \& M. Huisman}
\begin{document}
\maketitle
\def \bsl       {\symbol{92}}
\def \unsc      {\symbol{95}}

\newcommand{\msnote}[1]{{\sf {#1}}}

\newcommand{\calT}{{\cal T}}
\newcommand{\calS}{{\cal S}}
\newcommand{\calF}{{\cal F}}
\newcommand{\calA}{{\cal A}}
\newcommand{\calP}{{\cal P}}
\newcommand{\calQ}{{\cal Q}}
\newcommand{\calR}{{\cal R}}
\newcommand{\calX}{{\mathbf{X}}}
\newcommand{\calY}{{\mathbf{Y}}}
\newcommand{\calL}{{\cal L}}
\newcommand{\calH}{{\mathcal{H}}}
\newcommand{\calh}{{\mathit{H}}}
\newcommand{\cals}{{\mathit{S}}}
\newcommand{\icals}{{\mathit{S}_{T}^{\mathbf{i}}}}
\newcommand{\iscals}{{\mathit{S}^{\mathbf{i}}}}
\newcommand{\fcals}{{\mathit{S}_{T}^{\mathbf{f}}}}
\newcommand{\fscals}{{\mathit{S}_{\calo}^{\mathbf{f}}}}
\newcommand{\calo}{{\mathit{O}}}
\newcommand{\call}{{\mathit{L}}}
\newcommand{\calc}{{\mathit{C}}}
\newcommand{\shannon}{\ensuremath{\:\textit{Shan}\:}\xspace}
\newcommand{\renyi}{\ensuremath{\:\textit{R\'{e}nyi}\:}\xspace}
\newcommand{\Min}{\ensuremath{\:\textit{Min}\:}\xspace}

\newcommand{\mynote}[1]{{\it NOTE: #1}}
\newcommand{\calp}{{\mathcal{P}}}
\newcommand{\calf}{{\mathcal{F}}}
\newcommand{\PP}{{\mathbf{P}}}
\newcommand{\PVar}{{\mathit{Var}}}
\newcommand{\Lv}{{\mathit{L}}}
\newcommand{\Hv}{{\mathit{H}}}
\newcommand{\PVal}{{\mathit{W}}}
\newcommand{\distr}{{\mathcal{D}}}
\newcommand{\spt}{{\mathit{supp}}}
\newcommand{\dirac}{{\mathbf{1}}}
\newcommand{\NN}{{\mathbb{N}}}
\newcommand{\initstate}{I}
\newcommand{\valuation}{V}
\newcommand{\iparcomp}{{\mid \hspace*{-.1em} \mid}}
\newcommand{\HRule}{\rule{\linewidth}{0.3mm}}

\newcommand{\sem}[1]{\ensuremath{\llbracket{#1}\rrbracket}}

\newcommand{\assign}[2]{\ensuremath{{#1} \texttt{\,:=\,}} {#2}}
\newcommand{\comp}[2]{\ensuremath{{#1} \texttt{;}} {#2}}
\newcommand{\ifte}[3]{\ensuremath{\mathtt{if}\:\,\mathtt{(}{#1}\mathtt{)} \:\,\mathtt{then}\,\:
{#2}\, \:\mathtt{else}\,\: {#3}}}
\newcommand{\whileDo}[2]{\ensuremath{\mathtt{while}\: \mathtt{(}{#1}\mathtt{)\: do\:}
\: {#2}}}
\newcommand{\parcompprob}[2]{\ensuremath{{#1} \probchoice {#2}}}
\newcommand{\parcomp}[2]{\ensuremath{{#1} \mid \hspace*{-.2em} \mid {#2}}}
\newcommand{\wait}[1]{\ensuremath{\texttt{wait}\:\texttt{(}{#1}\texttt{)}}}

\newcommand{\myprooftree}[2]{\ensuremath{\frac{#1}{#2}}}

\newcommand{\act}[1]{\ensuremath{\langle #1 \rangle }}
\newcommand{\conf}[2]{\ensuremath{\langle #1, #2 \rangle }}
\newcommand{\confi}[1]{\ensuremath{\langle #1 \rangle }}
\newcommand{\confhist}[3]{\ensuremath{\langle #1, #2, #3 \rangle }}
\newcommand{\confknow}[3]{\ensuremath{\langle #1, #2, #3 \rangle }}
\newcommand{\semstep}[2]{\ensuremath{{#1} \rightarrow {#2}}}
\newcommand{\semstepp}[2]{\ensuremath{{#1} \rightarrow_p {#2}}}
\newcommand{\semstepm}[2]{\ensuremath{{#1} \rightarrow_m {#2}}}
\newcommand{\semstepextra}[2]{\ensuremath{{#1} \rightarrow_{\chi} {#2}}}
\newcommand{\semstepchi}[3]{\ensuremath{{#1} \stackrel{{#3}}{\rightarrow}{#2}}}
\newcommand{\semstepplus}[2]{\ensuremath{{#1} \rightarrow^+
{#2}}}
\newcommand{\semsteplab}[3]{\ensuremath{{#1} \xrightarrow{#3}{#2}}}

\newcommand{\hastrace}[2]{\ensuremath{{#1}\Downarrow{#2}}}
\newcommand{\hasinittrace}[2]{\ensuremath{{#1}\,\Downarrow_I\,{#2}}}

\newcommand{\defarrow}{\ensuremath{\overset{\mathit{def}} \Leftrightarrow}}
\newcommand{\defequal}{\ensuremath{\overset{\mathit{def}} =}}

\newcommand{\concat}[2]{\ensuremath{{#1}\,\hat{\ }\,{#2}}}
\newcommand{\conc}{\ensuremath{\,\hat{\ }\,}}

\newcommand{\rangeinteger}[1]{\ensuremath{\llbracket{#1}\rrbracket}}

\newcommand{\stutterequivstrong}[2]{\ensuremath{{#1}\,\cong_{L}\,{#2}}}
\newcommand{\stutterequiv}[2]{\ensuremath{{#1}\sim_{s}{#2}}}
\newcommand{\simupto}[2]{\ensuremath{\sim_{{#1},{#2}}}}
\newcommand{\simproupto}[2]{\ensuremath{\sim^p_{{#1},{#2}}}}
\newcommand{\traceupto}[2]{[{#1}, {#2}]}

\newcommand{\lhstore}[2]{\ensuremath{[{#1},{#2}]}}

\newcommand{\sumset}[1]{\ensuremath{\{ \hspace*{-3bp} \mid {#1} \mid \hspace*{-3bp} \} }}

\newcommand{\change}[2]{\ensuremath{\mathsf{c}_{{#1},{#2}}}\xspace}
\newcommand{\changex}[1]{\ensuremath{\mathsf{c}_{#1}}\xspace}
\newcommand{\changemod}[3]{\ensuremath{(\mathsf{c}_{{#1},{#2}})_{#3}}\xspace}
\newcommand{\changexmod}[2]{\ensuremath{(\mathsf{c}_{#1})_{#2}}\xspace}

\newcommand{\boxmod}[1]{\left[#1\right]}
\newcommand{\diamod}[1]{\left<#1\right>}
\newcommand{\wboxmod}[1]{\left[\hspace*{-.15em}\left[#1\right]\hspace*{-.15em}\right]}
\newcommand{\wdiamod}[1]{\left<\hspace*{-.2em}\left<#1\right>\hspace*{-.2em}\right>}
\newcommand{\hboxmod}[1]{\left[\hspace*{-.15em}\left[#1\right]\hspace*{-.15em}\right]_{H}}
\newcommand{\hdiamod}[1]{\left<\hspace*{-.2em}\left<#1\right>\hspace*{-.2em}\right>_{H}}

\newcommand{\pair}[2]{\ensuremath{({#1},{#2}})}
\newcommand{\triple}[3]{\ensuremath{({#1},{#2},{#3}})}

\newcommand{\stransition}{\overset{\epsilon} \Rightarrow}
\newcommand{\wtransition}[1]{\overset{{#1}} \Rightarrow}
\newcommand{\whhtransition}{\Rightarrow_{H'}}
\newcommand{\htransition}{\Rightarrow_H}
\newcommand{\whtransition}[1]{\overset{{#1}} \Rightarrow_H}

\newcommand{\lactions}{\ensuremath{\mathit{Act_L}}}
\newcommand{\hactions}{\ensuremath{\mathit{Act_H}}}

\newcommand{\ttt}{\mathsf{tt}}
\newcommand{\fff}{\mathsf{ff}}

\newcommand{\bp}{\phantom{(}}

\newcommand{\always}{\mathsf{always}}
\newcommand{\actionmod}[1]{(-)_#1^L}

\newcommand{\var}{\ensuremath{\mathit{AP}}\xspace}
\newcommand{\Rela}{\ensuremath{\mathcal{R}}\xspace}
\newcommand{\Rel}{\ensuremath{\mathsf{R}}\xspace}
\newcommand{\Trace}{\ensuremath{\mathit{Trace}}\xspace}
\newcommand{\Path}{\ensuremath{\mathit{Path}}\xspace}
\newcommand{\val}{\ensuremath{\mathit{Val}}\xspace}
\newcommand{\action}{\ensuremath{\mathit{Act}}\xspace}
\newcommand{\stmt}{\ensuremath{\mathit{Stmt}}\xspace}
\newcommand{\Config}{\ensuremath{\mathit{Conf}}\xspace}
\newcommand{\Com}{\ensuremath{\mathit{Com}}\xspace}
\newcommand{\Event}{\ensuremath{\Sigma}\xspace}
\newcommand{\Reach}{\ensuremath{\mathit{Reach}}\xspace}
\newcommand{\test}{\ensuremath{\mathit{test}}\xspace}
\newcommand{\escape}{\ensuremath{\mathit{escape}}\xspace}
\newcommand{\Vector}{\ensuremath{\mathit{Vec}}\xspace}
\newcommand{\re}{\ensuremath{\mathit{R}}\xspace}
\newcommand{\Ob}{\ensuremath{\mathit{ObS}}\xspace}
\newcommand{\SOb}{\ensuremath{\mathit{SObS}}\xspace}
\newcommand{\Obs}{\ensuremath{\mathit{SObS_{\delta}}}\xspace}
\newcommand{\Obsp}{\ensuremath{\mathit{SObS^*_{\delta}}}\xspace}
\newcommand{\ObE}{\ensuremath{\mathit{SObEnd}}\xspace}
\newcommand{\maxl}{\ensuremath{\mathit{maxL}}\xspace}
\newcommand{\abs}{\ensuremath{\mathit{Abs}}\xspace}
\newcommand{\Span}{\ensuremath{\mathit{Span}}\xspace}
\newcommand{\Prob}{\ensuremath{\mathcal{P}\mathit{r}}\xspace}
\newcommand{\Store}{\ensuremath{\mathit{Store}}\xspace}
\newcommand{\storeinit}{\ensuremath{\mathit{Store_{init}}\xspace}}
\newcommand{\term}{\ensuremath{\mathit{Term}}\xspace}
\newcommand{\trueC}{\ensuremath{\mathtt{true}}\xspace}
\newcommand{\falseC}{\ensuremath{\mathtt{false}}\xspace}
\newcommand{\hist}{\ensuremath{\mathcal{H}}\xspace}
\newcommand{\Hist}{\ensuremath{\mathit{Hist}}\xspace}
\newcommand{\seq}{\ensuremath{\mathit{Seq}}\xspace}
\newcommand{\runh}{\ensuremath{\mathit{Run}_{H}}\xspace}
\newcommand{\confacc}[1]{\ensuremath{\mathit{conf}_{\!#1}}\xspace}
\newcommand{\current}{\ensuremath{\mathit{current}}\xspace}
\newcommand{\old}{\ensuremath{\mathit{old}}\xspace}
\newcommand{\trans}{\ensuremath{(c_0,\delta)}\xspace}
\newcommand{\statel}{\ensuremath{S}\xspace}
\newcommand{\pa}{\ensuremath{\mathcal{B}}\xspace}
\newcommand{\ipa}{\ensuremath{\mathcal{B}_{0}(l)}\xspace}
\newcommand{\replace}{\ensuremath{\mathit{Replace}}\xspace}
\newcommand{\bottom}{\ensuremath{\mathit{Bottom}}\xspace}
\newcommand{\tops}{\ensuremath{\mathit{top}}\xspace}
\newcommand{\pres}{\ensuremath{\mathit{Pre}}\xspace}
\newcommand{\bad}{\ensuremath{\mathit{Bad}}\xspace}
\newcommand{\Stable}{\ensuremath{\mathit{Stable}}\xspace}
\newcommand{\findSpli}{\ensuremath{\mathit{find \unsc Splitter}}\xspace}
\newcommand{\inloop}{\ensuremath{\mathit{Loop}}\xspace}
\newcommand{\final}{\ensuremath{F}\xspace}
\newcommand{\SSObsDet}{\textsf{SSOD}\xspace}
\newcommand{\SSProObsDet}{\textsf{SSPOD}\xspace}
\newcommand{\Fair}{\ensuremath{\mathit{Fair}}\xspace}

\newcommand{\stf}{\ensuremath{\mathit{sf}}\xspace}
\newcommand{\dsf}{\ensuremath{\mathit{dsf}}\xspace}
\newcommand{\stut}{\ensuremath{\mathit{stut}}\xspace}
\newcommand{\St}{\ensuremath{\mathit{Q}}\xspace}
\newcommand{\suc}{\ensuremath{\mathit{succ}}\xspace}
\newcommand{\pred}{\ensuremath{\mathit{pred}}\xspace}
\newcommand{\queue}{\ensuremath{\mathit{Q}}\xspace}
\newcommand{\diverqueue}{\ensuremath{\mathit{Non \unsc Divergent \unsc Q }}\xspace}
\newcommand{\DiverQ}{\ensuremath{\mathit{Divergent \unsc Q }}\xspace}
\newcommand{\parent}{\ensuremath{\mathit{parent}}\xspace}
\newcommand{\mapp}{\ensuremath{\mathit{map}}\xspace}
\newcommand{\countd}{\ensuremath{\mathit{count}}\xspace}
\newcommand{\diver}{\ensuremath{\mathit{divergence}}\xspace}
\newcommand{\witness}{\ensuremath{\mathit{witness \unsc trace}}\xspace}
\newcommand{\witnesslength}{\ensuremath{\mathit{witness \unsc length}}\xspace}
\newcommand{\witnessindex}{\ensuremath{\mathit{witness \unsc index}}\xspace}
\newcommand{\witnesscurrent}{\ensuremath{\mathit{witness \unsc current}}\xspace}
\newcommand{\loopstart}{\ensuremath{\mathit{loop \unsc start}}\xspace}
\newcommand{\Stm}{\ensuremath{\mathit{Inv}}\xspace}
\newcommand{\fst}{\ensuremath{\mathit{Fst}}\xspace}
\newcommand{\scd}{\ensuremath{\mathit{Scd}}\xspace}
\newcommand{\stack}{\ensuremath{\mathit{Stack}}\xspace}
\newcommand{\position}{\ensuremath{\mathit{position}}\xspace}
\newcommand{\indexwitness}{\ensuremath{\mathit{index}}\xspace}
\newcommand{\corres}{\ensuremath{\mathit{{candidate \unsc fbis}}}\xspace}
\newcommand{\simlabel}{\ensuremath{\sim_{\valuation}}\xspace}
\newcommand{\abnormal}{\ensuremath{\mathit{{abnormal \unsc state}}}\xspace}
\newcommand{\preabnormal}{\ensuremath{\mathit{{pre \unsc abnormal \unsc state}}}\xspace}
\newcommand{\probchoice}{\ensuremath{\oplus_p}\xspace}

\newcommand{\size}{\ensuremath{\mathit{size}}\xspace}
\newcommand{\flag}{\ensuremath{\mathit{flag}}\xspace}
\newcommand{\lab}{\ensuremath{\mathit{Map}}\xspace}
\newcommand{\Label}{\ensuremath{\mathit{u}}\xspace}
\newcommand{\Visit}{\ensuremath{\mathit{Visit}}\xspace}
\newcommand{\true}{\ensuremath{\mathit{true}}\xspace}
\newcommand{\false}{\ensuremath{\mathit{false}}\xspace}
\newcommand{\continue}{\ensuremath{\mathit{continue}}\xspace}
\newcommand{\Lab}{\ensuremath{\mathit{lab}}\xspace}
\newcommand{\terminated}{\ensuremath{\mathit{final}}\xspace}
\newcommand{\nextLab}{\ensuremath{\mathit{next \unsc lab}}\xspace}
\newcommand{\potentialLab}{\ensuremath{\mathit{potential \unsc map}}\xspace}
\newcommand{\undefined}{\ensuremath{\mathit{undefined}}\xspace}
\newcommand{\rootv}{\ensuremath{\mathit{root}}\xspace}
\newcommand{\init}{\ensuremath{\mathit{init \unsc state}}\xspace}
\newcommand{\initlabel}{\ensuremath{\mathit{init \unsc label}}\xspace}
\newcommand{\pvalue}{\ensuremath{\mathit{parent \unsc value}}\xspace}
\newcommand{\sameblock}{\ensuremath{\mathit{same \unsc block \unsc flag}}\xspace}
\newcommand{\trace}{\ensuremath{\mathit{trace}}\xspace}
\newcommand{\lgth}{\ensuremath{\mathit{length}}\xspace}
\newcommand{\unmatched}{\ensuremath{\mathit{Un\unsc matched}}\xspace}
\newcommand{\source}{\ensuremath{\mathit{source}}\xspace}
\newcommand{\dest}{\ensuremath{\mathit{dest}}\xspace}
\newcommand{\rel}{\ensuremath{\mathit{R_{nd}}}\xspace}
\newcommand{\States}{\ensuremath{\mathit{States}}\xspace}
\newcommand{\vsucc}{\ensuremath{\mathit{Valid\unsc Succ}}\xspace}
\newcommand{\ivsucc}{\ensuremath{\mathit{Invalid\unsc Succ}}\xspace}

\newcommand{\composedstate}{\ensuremath{\mathfrak{c}}\xspace}
\newcommand{\add}{\ensuremath{\mathit{add}}\xspace}
\newcommand{\first}{\ensuremath{\mathit{first}}\xspace}
\newcommand{\length}{\ensuremath{\mathit{length}}\xspace}
\newcommand{\otherlength}{\ensuremath{\mathit{lgth\unsc othr \unsc trc}}\xspace}
\newcommand{\last}{\ensuremath{\mathit{last}}\xspace}
\newcommand{\remove}{\ensuremath{\mathit{remove}}\xspace}

\newcommand{\leading}{\ensuremath{\mathit{lead}}\xspace}
\newcommand{\extra}{\ensuremath{\mathit{extra}}\xspace}
\newcommand{\nrchanges}[1]{\ensuremath{\mathit{nr\unsc ch}^{#1}}\xspace}
\newcommand{\nrchangesp}[1]{\ensuremath{{\mathit{nr\unsc ch}^{#1}}'}\xspace}
\newcommand{\firstchange}[2]{\ensuremath{\mathit{first\unsc change}_{#2}^{#1}}\xspace}
\newcommand{\secondchange}[2]{\ensuremath{\mathit{second\unsc change}_{#2}^{#1}}\xspace}
\newcommand{\nrchangesless}[3]{\ensuremath{\mathit{nr\unsc ch_{#3}}^{#1} < \mathit{nr\unsc ch}_{#3}^{#2} }\xspace}
\newcommand{\nc}{\ensuremath{\mathsf{nc}}\xspace}
\newcommand{\nochange}{nochange}
\newcommand{\pretrace}{\ensuremath{\sigma} \xspace}
\newcommand{\scheduler}{\ensuremath{\delta} \xspace}

\newcommand{\fm}{\ensuremath{\mathit{FM}^{\phi}(K)}}
\newcommand{\tk}{\ensuremath{\mathit{TK}}\xspace}

\newcommand{\ie}{{\itshape i.e.}, }
\newcommand{\eg}{{\itshape e.g.}, }
\newcommand{\cf}{{\itshape cf.}\ }
\newcommand{\wrt}{{\itshape w.r.t.}\xspace}
\newcommand{\etal}{{\itshape et al.}\xspace}





\newcounter{programcounter}
\newenvironment{program}[1][]%
{\refstepcounter{programcounter}{#1}{\small}{}}

\newcommand{\dom}{\ensuremath{\mathsf{dom}}\xspace}

\newcommand{\prog}{\ensuremath{\mathit{prog}}\xspace}
\newcommand{\store}{\ensuremath{\mathit{store}}\xspace}
\newcommand{\histf}{\ensuremath{\mathsf{hist}}\xspace}
\newcommand{\mustore}{\ensuremath{\mathsf{store}}\xspace}
\newcommand{\termpred}[1]{\ensuremath{\mathsf{term}(#1)}\xspace}
\newcommand{\deadlock}{\ensuremath{\mathsf{deadlock}}\xspace}
\newcommand{\readV}{\ensuremath{\mathsf{readV}}\xspace}
\newcommand{\writeV}{\ensuremath{\mathsf{writeV}}\xspace}

\newcommand{\eliminate}{\ensuremath{\upharpoonright}}
\newcommand{\filter}{\ensuremath{\mathsf{filter}}\xspace}
\newcommand{\map}{\ensuremath{\mathit{map}}\xspace}
\newcommand{\reach}{\ensuremath{\mathsf{reach}}\xspace}

\newcommand{\havoc}{\ensuremath{\mathit{HH}}}

\newcommand{\restrictto}[2]{\ensuremath{#1\phantom{'}\!}_{|_{#2}}}

\newcommand{\tick}{\ensuremath{\surd}}
\newcommand{\cross}{\ensuremath{{\large \times}}}

\newcommand{\stoch}[3]{\ensuremath{\mathcal{M}^{#1}_{#2,#3}}}
\newcommand{\p}[3]{\ensuremath{\mathcal{P}_{#1}({#2} \rightarrow {#3})}}
\newcommand{\pc}[2]{\ensuremath{p_{#1}({#2})}}

\newcommand{\Dist}[1]{\ensuremath{\mathit{Dist}({#1})}}
\newcommand{\Distzero}[1]{\ensuremath{\mathit{Dist}_0({#1})}}
\newcommand{\dist}{\ensuremath{\mathsf{dist}}}

\newcommand{\nat}{\ensuremath{\mathbb{N}}}

\newcommand{\mimic}{\ensuremath{\mathit{mimic}}}

\newcommand{\actionod}{\ensuremath{\mathit{Act}}}

\newcommand{\pre}[2]{\ensuremath{{#1}_{\ll{#2}}}}
\newcommand{\sub}[2]{\ensuremath{{#1}_{\gg{#2}}}}

\newsavebox{\mysquare}
\newlength\mytextsize
\newcommand{\textWithLine}[2]{%
\rule[.25\baselineskip]{\textwidth}{0.25mm}\\
\settowidth{\mytextsize}{#1}
\savebox{\mysquare}{%
     \hspace{#2}%
     \raisebox{\baselineskip}[.25\baselineskip][.25\baselineskip]{%
       \textcolor{white}{%
     \rule{\mytextsize}{.6em}%
       }%
       \hspace{-\mytextsize}#1
     }%
}
\usebox{\mysquare}\vspace{-\baselineskip}\\
}

\newcommand{\rectangle}{\scalebox{0.8}{[\!]}}
\begin{abstract}
Quantitative theories of information flow give us an approach to relax the absolute confidentiality properties that are difficult to satisfy for many practical programs. The classical information-theoretic approaches for sequential programs, where the program is modeled as a communication channel with only input and output, and the measure of leakage is based on the notions of \emph{initial} uncertainty and \emph{remaining} uncertainty after observing the final outcomes, are not suitable to multi-threaded programs. Besides, the information-theoretic approaches have been also shown to conflict with each other when comparing programs. Reasoning about the exposed information flow of multi-threaded programs is more complicated, since the outcomes of such programs depend on the scheduler policy, and the leakages in intermediate states also contribute to the overall leakage of the program. 

This paper proposes a novel model of quantitative analysis for multi-threaded programs that also takes into account the effect of observables in intermediate states along the trace. We define a notion of the leakage of a program trace. Given the fact that the execution of a multi-threaded program is typically described by a set of traces, the leakage of a program under a specific scheduler is computed as the expected value of the leakages of all possible traces. Examples are given to compare our approach with the existing approaches.
\end{abstract}
\section{Introduction}\label{Intro}
\paragraph{Qualitative and Quantitative Security Properties.}
Qualitative security properties, such as \emph{noninterference} \cite{GoguenM82} and \emph{observational determinism} \cite{ZM,Huisman11}, are appropriate for applications like Internet banking, e-commerce, and medical information systems, where private data, like credit card details, and medical records, need strict protection. The key idea of such absolutely confidential properties is that secret information should not be derivable from public data. These properties prohibit any information flow from a high security level to a low level\footnote{For simplicity, throughout this paper, we consider a simple two-point security lattice, where the data is divided into two disjoint subsets, of private (high) and public (low) security levels, respectively.}. 
For example, the program $\mathtt{if \: (\cals>0)\:\:then \:\: \calo:=0\:\:else\:\:\calo:=1}$, where $\cals$ is a private variable and $\calo$ is a public variable, is rejected by qualitative security properties, since we can learn information about $\cals$ from the value of $\calo$.
  
However, for many applications in which we want or need to reveal information that depends on private data, these absolutely confidential properties are not appropriate. A program that contains a leakage is rejected even when the leakage is \emph{unavoidable}. An typical example is a \emph{password checker}, where an attacker tries a string to guess the password. Even when the attacker makes a wrong guess, a small amount of secret information is still leaked, i.e., the attacker is able to derive that the password is not the string that he has just entered. In this example, the leakage is unavoidable.

Therefore, an alternative approach is to relax the definitions of absolute confidentiality by quantifying information flow and determining how much secret information is being leaked, i.e., expressing the amount of leakage in quantitative terms. A quantitative security policy offers a method to compute \emph{bounds} on how much information is leaked. This information can be used to decide whether we can tolerate \emph{minor} leakages. Quantifying information flow also provides a way to judge whether an application leaks more information than another, although they may both be insecure. A quantitative theory thus can be seen a generalization of a theory of absolute confidentiality.

\paragraph{Quantitative Security Analysis for Sequential Programs.} 
The foundation of quantitative analysis of information flow for sequential programs is the scenario that a program is considered as a channel in the information-theoretic sense, where the secret $\cals$ is the input and the observables $\calo$ are the output. 
An attacker, by observing $\calo$, might be able to derive information about $\cals$. 

Quantitative security analysis studies how much information about $\cals$ an attacker might learn from the output $\calo$.
The analysis uses the notion of entropy. The entropy of a random variable expresses the \emph{uncertainty} of an attacker about its value, i.e., how \emph{difficult} it is for an attacker to discover its value. 
The leakage of a program is then defined as the difference between the secret's initial uncertainty, i.e., the uncertainty of the attacker about private data before executing the program, and the secret's remaining uncertainty, i.e., the uncertainty of the attacker after observing the program's public outcomes, i.e.,

\begin{center}
Information leakage = Initial uncertainty - Remaining uncertainty.
\end{center}
In general, by observing the outcomes of the program, the attacker gains more knowledge about the initial private data. Thus, this reduces the initial uncertainty and causes the information leakage.

However, the existing approaches for sequential programs do not agree on a unique measure to quantify information flow. Past works
have proposed several entropy measures to compute the program's leakage, i.e., \emph{Shannon entropy}, \emph{R\'{e}nyi's min-entropy} and \emph{Guessing entropy} (see \cite{Alvim11} for more details).
Since we expect that the right notion of entropy should make the value of the computed leakage non-negative,  
Guessing entropy is not suitable to quantify information flow of our interested programs in which the output is non-deterministic and probabilistic. Guessing entropy only guarantees the non-negativeness property of leakage for deterministic programs, i.e., the programs such that for each input, a unique output is produced. Therefore, for probabilistic systems, Shannon-entropy and R\'{e}nyi's min-entropy are more appropriate. However, these measures might be in conflict, i.e.,  measures based on Shannon-entropy judge one program more dangerous than the other while min-entropy measures give an opposite result. Thus, it seems that no single measure is likely to be appropriate in all cases \cite{Alvim12}.


\paragraph{Quantitative Security Analysis for Multi-threaded Programs.}
With the trend of parallel systems, and the emergence of multi-core computing, multi-threading is becoming more and more standard. However, the question which measure, i.e., Shannon entropy  or min entropy, is appropriate to quantify information flow of multi-threaded programs still needs an answer. Besides,
the existing models of quantitative analysis are not appropriate 
for multi-threaded programs, since they consider only the input and the final outcomes of a program. For multi-threaded programs, due to the interactions between threads and the exchange of intermediate results, intermediate states should also be taken into account (see \cite{ZM,Huisman11} for more details). Thus, new methods have to be developed for an observational model where an attacker can access the full code of the program, and observe the traces of public data.

To obtain a suitable model of analyzing quantitatively information flow for multi-threaded programs, we need to
\begin{inparaenum}[(i)]
	\item consider how the public values in the intermediate states also help an attacker to reduce his initial uncertainty about private data. Consequently, we need to define the leakage of an execution trace, i.e., the leakage given by a sequence of publicly observable data obtained
during the execution of the program, and
	\item take into account the effect of schedulers on the overall leakage of the programs, since the outcomes of a multi-threaded program depend on the scheduling policy. Thus, we need to consider how the distribution of traces affects the overall leakage, i.e., the execution of a multi-threaded program always results in a set of traces. 
\end{inparaenum}

This paper proposes a novel approach of quantitative analysis for multi-threaded programs. We model the execution of a multi-threaded program under the control of a probabilistic scheduler by a probabilistic state transition system. The probabilities of the transitions are decided by the scheduler that is used to deploy the program. States denote the probability distributions of private data $\cals$. The distribution of $\cals$ changes from state to state along a trace, depending on the public values in the states and the program commands that result in such observables. This idea is sketched by the following example.
Initially, given that an attacker knows that the true value of $\cals$ is in the set $\{0,1,2,3\}$, his initial knowledge might be expressed by a uniform distribution of $\cals$, i.e., $\{0 \mapsto \frac{1}{4}, 1 \mapsto \frac{1}{4}, 2 \mapsto \frac{1}{4}, 3 \mapsto \frac{1}{4}\}$. Two program commands $\calo:=\cals/2$ and $\calo:=\cals \mod 2$ might result in the same $\calo$, e.g., $\calo=1$, but if the result is from $\calo:=\cals/2$, the updated distribution is $\{0 \mapsto 0, 1 \mapsto 0, 2 \mapsto \frac{1}{2}, 3 \mapsto \frac{1}{2} \}$, otherwise, it is $\{0 \mapsto 0, 1 \mapsto \frac{1}{2}, 2 \mapsto 0, 3 \mapsto \frac{1}{2}\}$. 

The program can be seen a distribution transformer. During the program execution, the distribution of private data transforms from the initial distribution to the final ones in traces. The distributions of private data at the initial and final states of a trace can be used to present the \emph{initial} uncertainty of the attacker about secret information, and his \emph{final} uncertainty, after observing the sequence of public data along the trace, respectively. Based on these notions,
we define the leakage of an execution trace as the difference between the \emph{initial uncertainty} and the \emph{final uncertainty}. 


The quantitative term to denote uncertainty relates much to the model of attacker \cite{Alvim11}, i.e., an attacker can guess secret information by multiple tries, or just by one try. 
In our approach, we follow the \emph{one-try threat} model, which is more suitable to many security situations i.e., the system will trigger an alarm when an attacker makes a wrong guess. Based on this threat model, we denote the initial and final uncertainty of the attacker by R\'{e}nyi's min-entropies of the initial and final distributions of private data, respectively. Notice that in our approach, instead of the notion of \emph{remaining} uncertainty as in classical approaches, we use the notion of \emph{final} uncertainty. 
While remaining uncertainty depends only on the outcomes of the program, our notion of final uncertainty depends on the observables along the trace, and also on the program commands that result in such observables. 



Given a multi-threaded program, a scheduler, the execution of the program under the control of the scheduler always results in a set of traces. The leakage of the program is then computed as the \emph{expected} value of the leakages of traces. Via a case study (Section~\ref{ananlysisprocess}), 
we demonstrate how the leakage of a multi-threaded programs is measured. We also compare our approach with the existing quantitative analysis models. We believe that our approach gives a more accurate way to study quantitatively the security property of multi-threaded programs.

\paragraph{{\bf Organization of the Paper.}}
Section~\ref{Preliminaries} presents the preliminaries. Then,
Section~\ref{ClassicalApp} discusses why the classical approaches are not suitable to quantify information flow of multi-threaded programs.
Section~\ref{OurApp} presents our quantitative analysis model, a case study, and also the comparison with the existing approaches.
Section~\ref{Relatedwork} discusses related work.
Section~\ref{conclusion} concludes, and discusses future work.
\vspace*{-1em}
\section{Preliminaries}\label{Preliminaries}
\subsection{Probabilistic Distribution}
Let $X$ and $Y$ denote two discrete random variables with carriers $\mathbf{X}=\{x_1,\ldots,x_n\}$, $\mathbf{Y}=\{y_1,\ldots,y_m\}$, respectively. A \emph{probability distribution} $p_X(\cdot)$ over a set $\mathbf{X}$ is a function $p_X\: : \mathbf{X} \rightarrow [0,1]$, such that the sum of the probabilities of all elements over $\mathbf{X}$ is 1, i.e., $\sum_{x \in \mathbf{X}}p_X(x)=1$. If $\mathbf{X}$ is uncountable, then $\sum_{x \in \mathbf{X}}p_X(x)=1$ implies that $p_X(x)>0$ for countably many $x \in \mathbf{X}$. 

We use $p_{X|Y}(x|y)$ to denote the \emph{conditional} probability that $X=x$ when $Y=y$. We call $p_X(\cdot)$ a \emph{priori} distribution of $X$, and $p_{X|Y}(\cdot|\cdot)$ a \emph{posteriori} distribution. Thus, $p(x)$ and $p(x|y)$ are a \emph{priori} and a \emph{posteriori} probabilities of $x$. 

\subsection{Shannon Entropy}
\begin{definition}
The Shannon entropy of a random variable $X$ is defined as \cite{Alvim11},
\[\calH_{\shannon}(X)=-\sum_{x \in \calX} p(x) \log p(x),\]
where the base of the logarithm is set to $2$.
\end{definition}

\begin{definition}
The conditional Shannon entropy of a random variable $X$ given $Y$ is \cite{Alvim11},
\[\calH_{\shannon}(X|Y)=\sum_{ y \in \calY} p(y) \calH_{\shannon}(X|Y=y),\]
where $\calH_{\shannon} (X|Y=y)= -\sum_{ x \in \calX} p(x|y)\log p(x|y)$.
\end{definition}

It is possible to prove that $0\leq \calH_{\shannon}(X|Y)\leq \calH_{\shannon}(X)$. The minimum value of $\calH_{\shannon}(X|Y)$ is $0$, if $X$ is completely determined by $Y$. The maximum value of $\calH_{\shannon}(X|Y)$ is $\calH_{\shannon}(X)$, when $X$ and $Y$ are independent.

\subsection{Min-Entropy}
\begin{definition}
The R\'{e}nyi's min-entropy of a random variable $X$ is defined as \cite{Smith09},
\[\calH_{\renyi}(X)= - \log\:\max_{x \in \calX} p(x),\]
\end{definition}
Notice that Shannon entropy and R\'{e}nyi's min-entropy coincide on uniform distributions.

R\'{e}nyi did not define the notion of conditional min-entropy, and there is no unique definition of this notion. Smith \cite{Smith09} considers the following definition.
\begin{definition}
The conditional min-entropy of a random variable $X$ given $Y$ is,
\[\calH_{\Min}(X|Y)= - \log \sum_{ y \in \calY}  p(y) \cdot \max_{x \in \calX} p(x|y).\]
\end{definition}

\section{Classical Models of Quantitative Information Flow}\label{ClassicalApp}
This section presents classical models of quantitative security analysis, and discusses why they are not suitable to quantify information leakage of multi-threaded programs. 

\subsection{Information Leakage in the Classical Approaches}
Existing works \cite{Moskowitz03,Clark05,Chatzikokolakis06,Malacaria08,Malacaria10,Zhu05,Smith09,Andres10} propose to use 
information theory as a setting to model information flow. A program is seen a standard input-output model with the secret $\cals$ as the input and the public outcomes $\calo$ as the output. 
Let $\calH(\cals)$ denote the initial uncertainty of an attacker about secret information, and $\calH(\cals|\calo)$ denote the uncertainty after the programs has been executed and the public outcomes are observed.
The leakage of the program is given by,

\begin{center}
$\calL(P) = \calH(\cals) - \calH(\cals|\calo)$.
\end{center}
where $\calL(P)$ denotes the leakage of $P$; $\calH$ might be either Shannon entropy or min-entropy. 


\subsection{Basic Settings for Leakage} \label{basicsettings}
To argue why a program is considered more dangerous than the another, we need to setup some basic settings for the discussion. 


First, we consider the \emph{one-try threat model}, i.e., the attacker is allowed to guess the value of $\cals$ by only one try.
Secondly, it is often the case that in practice, the attacker only knows the possible values of private data, but he does not know which value is likely to be the correct one. Thus, the attacker's initial knowledge about the secret is denoted by the uniform distribution on its possible values. In other words, we assume that initially, the attacker knows nothing but the uniform prior distribution of private data. 

Finally, we also assume that systems are two-point security lattice, and data in the same category, i.e., high or low security level, are indistinguishable in \emph{security meaning}. Thus, a system that leaks the last 9 bits of private data is considered just as dangerous as a system that leaks the first 9 bits. 

\subsection{Classical Measures might be Counter-intuitive and Conflict}
Many authors \cite{Moskowitz03,Clark05,Chatzikokolakis06,Malacaria08,Malacaria10,Zhu05} measure leakage using Shannon entropy. However, Smith \cite{Smith09} shows that in the context of the one-try threat model, measures based on Shannon entropy do not always result in a very good operational security guarantee. In particular, Smith \cite{Smith09} shows that Shannon-entropy measures might be counter-intuitive. He shows two programs $P$ and $P'$ such that by intuitive understanding, $P$ leaks more information than $P'$, but Shannon-entropy measures yield a counter-intuitive result, i.e., $\calL_{\shannon}(P)<\calL_{\shannon}(P')$, where $\calL_{\shannon}()$ denotes the leakage given by Shannon-entropy measure.
For example, consider the two following programs $P_1$ and $P_2$, from \cite{Smith09}.
\begin{example}[Program $P_1$]
\[\mathtt{if \: (\cals\!\!\!\!\!\! \mod 8 = 0)\:\:then \:\: \calo:=\cals\:\:else\:\:\calo:=1;}\]
\label{example1} 
\end{example}
Basically, $P_1$ copies $\cals$ to $\calo$ when $\cals$ is a multiple of $8$, otherwise, it sets $\calo$ to $1$. Assume that $\cals$ is a 64-bit unsigned integer, $0\leq \cals < 2^{64}$. According to Shannon-entropy measure \cite{Smith09}, the leakage of $P_1$ is 
$\calL_{\shannon}(P_1) =8.17$.

\begin{example}[Program $P_2$]
\[\mathtt{\calo:=\cals \:\&\: (0\ldots 0.111.111.111)_b;}\]
where $\mathtt{(0\ldots 0.111.111.111)_b}$ is the 64-bit binary number such that the first 55 bits are $0$ and the last 9 bits are $1$.\label{example2} 
\end{example}
This program simply copies the last 9 bits of $\cals$ into $\calo$. Shannon-entropy measure gives 
$\calL_{\shannon}(P_2) =9$.

In $P_1$, whenever the public outcome $\calo \neq 1$, the attacker obtains the secret $\cals$ completely. Thus, the expected probability of guessing $\cals$ by one try is greater than $\frac{1}{8}$. In $P_2$, for any value of $\calo$, the probability of guessing $\cals$ by one try is $2^{-55}$, since the first $55$ bits of $\cals$ are still unknown. It means that, in the one-try threat model, intuitively, $P_1$ is considered more dangerous than $P_2$. However, the measures based on Shannon entropy judge $P_2$ worse than $P_1$. 

For this reason, Smith develops an alternative theory of quantitative information flow based on min-entropy \cite{Smith09}.
Smith defines uncertainty in terms of the \emph{vulnerability} of $\cals$ to be guessed in one try. The vulnerability of a random variable $X$ is the maximum of the probabilities of the values of $X$. This approach seems match the intuitive idea of the one-try threat model, i.e., the attacker always chooses the value with the maximum probability. 

However, min-entropy measures might still result in counter-intuitive values of leakage. Consider the two following programs $P_3$ and $P_4$, from \cite{Smith09,Zhu10}.
\begin{example}[Program $P_3$: Password Checker]
\[\mathtt{if \: (\cals=\call)\:\:then \:\: \calo:=1\:\:else\:\:\calo:=0;}\]
where $\cals$ denotes the password, $\call$ the string entered by the attacker, and $\calo$ the observable answer, i.e., right or wrong.\label{example3} 
\end{example}

\begin{example}[Program $P_4$: Binary Search]
\[\mathtt{if \: (\cals\geq\call)\:\:then \:\: \calo:=1\:\:else\:\:\calo:=0;}\]
where $\call=|\cals|/2$ is a program parameter. \label{example4}
\end{example}

Assume that $\cals$ is a uniformly-distributed unsigned integer. The measures based on min-entropy do not distinguish between $P_3$ and $P_4$ by always judging that their leakages are the same, i.e., $\calL_{\Min}(P_3) = \calL_{\Min}(P_4)=1$. However,
if $|\cals|$ is large, the probability that $\cals=\call$ becomes so low in $P_3$, i.e., $p(\calo=1)=\frac{1}{|\cals|}\approx 0$. Thus, intuitively, $P_3$ leaks almost nothing, since the probability of one-try guessing $\cals$ after observing the outcome is $\frac{1}{|\cals|-1} \approx \frac{1}{|\cals|}$. Program $P_4$ always leaks 1 bit of information, since the probability of guessing $\cals$ after observing the outcome is $\frac{2}{|\cals|}$. Thus, program $P_4$ should be judged more dangerous than $P_3$. In his paper \cite{Smith09}, Smith also admits that $P_3$ and $P_4$ should be treated differently. It is trivial that the uncertainty of the password guessing decreases slowly, while in binary search, the uncertainty of the secret decreases very rapidly.

Notice that for this example, Shannon-entropy measures give $\calL_{\shannon}(P_3)=0$ and $\calL_{\shannon}(P_4)=1$, that match the intuition. Therefore, there is growing appreciation that no single leakage measure is likely to be appropriate in all cases. This puts a question which measure is more suitable to quantify information leakage of multi-threaded programs.

%
%
%

\subsection{Leakages in Intermediate States}
Classical approaches of quantitative security analysis were based on the input-output model. However, for multi-threaded programs, due to the interactions between threads and the exchange of intermediate results, one should also consider the leakages in intermediate states along the traces. Consider the following example where $\cals$ is a 3-bit binary number.

\begin{example} Given that $(100)_b$ and $(011)_b$ are the binary forms of $4$ and $3$, respectively.
\[
\begin{array}{rl}
& \calo:=0;  \\
& \calo:=\cals \:\&\: (100)_b;\\
& \calo:=\cals \:\&\: (011)_b;
\end{array}
\]
\label{interstate1}  
\end{example}
Let $(s_3 s_2 s_1)_b$ denote the binary form of $\cals$. The existing input-output models judge that this program leaks $2$ bits of private data in the output, i.e., $\calo=(s_3 s_2 s_1)_b \:\&\: (011)_b = (0 s_2 s_1)_b$. However, due to the leakage in the intermediate state, i.e., $\calo=(s_3 s_2 s_1)_b \:\&\: (100)_b = (s_3 00)_b$, the attacker obtains the full secret value.

However, notice that leakages in intermediate states do not always contribute to the overall leakage of a trace, as in the following example,
\begin{example}
\[
\begin{array}{rl}
& \calo:=0;  \\
& \calo:=\cals \:\&\: (001)_b;\\
& \calo:=\cals \:\&\: (011)_b;
\end{array}
\]\label{interstate2}
\end{example}  

The overall leakage of this program trace is only $2$ bits, since the leakage in the intermediate state, i.e., the last bit $s_1$, is included in the leakage in the output. This example shows that leakage of a program trace is not simply the sum of leakages of transition steps along the trace, as in the approach of Chen et al. \cite{Chen09} (We discuss this approach more in the Related Work). 

\subsection{Scheduler's Effect}
The classical quantitative analysis of information flow do not take into account the effect of schedulers.
Since the set of possible traces of a multi-threaded program depends on the scheduler that is used to execute the
program, schedulers cannot be ignored when quantifying information flow.
Consider the following example,
\begin{example} Given that $\big|\!\big|$ is the parallel operator, and $\cals$ is a 2-bit secret information,
\[\calo:=\cals/2\:\big|\!\big|\:\calo:=\cals\!\!\!\!\!\mod 2.\]
\label{trace}                   
\end{example}

Execute this program with a uniform scheduler, i.e., a scheduler that picks threads with the same probability in a time slot. Since $\cals$ is a uniform  2-bit data, there are 4 possible traces with the same probability of occurrence, i.e., $\{00,01,10,11\}$. If the obtained trace is either $00$ or $11$, the attacker can conclude for sure that the value of $\cals$ is $0$, or $3$, respectively. However, if the trace is $01$ or $10$, he is only able to derive that $\cals$ is either $1$ or $2$. Thus, with this scheduler, the secret is not leaked totally.

However, when the attacker chooses a scheduler which always executes $\calo:=\cals/2$ first, he also obtains 4 possible traces $\{00,01,10,11\}$ with the same probability of occurrence, but in this case, he can always derive the value of $\cals$ correctly.


%

\section{Quantitative Security Analysis for Multi-threaded Programs}\label{OurApp}
\subsection{Probabilistic Kripke Structures}
We consider probabilistic Kripke structures (PKS) that can be used to model the semantics of probabilistic programs in a standard way \cite{Gurfinkel06,Ngo13}. PKSs are like standard Kripke structures \cite{Kripke1963}, except that each state transition $c \rightarrow \mu$ leads to a probability distribution $\mu$ over the next states, i.e., the probability to end up in state $c'$ is $\mu(c')$ ($c,c' \in \mbox{a set of states } \Config$). Each state may enable several probabilistic transitions, modeling different execution orders to be determined by a scheduler. 

We assume that the set of variables of the program $\PVar$ is partitioned into a set of low variables $\calo$ and a set of high variables $\cals$, i.e., $\PVar = \calo \cup \cals$, with $\calo \cap \cals = \emptyset$. To aim for simplicity and clarity, rather than full generality, following \cite{Smith09}, our development therefore restricts to 
programs with only one high and one low variables. Our goal is to quantify how much information about $\cals$ is deduced by an attacker who can observe the set of execution traces of $\calo$. This restriction aims to demonstrate the core idea of the analysis process. However, the analysis can adapt to more complex multi-threaded programs easily after some trivial modifications.

Our PKSs label states with the distributions of $\cals$. Thus, each state $c$ is labeled by a labeling function $V :\:\Config \rightarrow \distr(\cals)$  that assigns a distribution $p_{\cals} \in \distr(\cals)$ to each $c \in \Config$. 

\begin{definition} [Probabilistic Kripke structure]
A \emph{probabilistic Kripke structure} $\calA$ is a tuple
$\langle \Config, \initstate, \allowbreak \PVar, \valuation, \rightarrow \rangle$ 
 consisting of
\begin{inparaenum}[(i)]
	\item a set $\Config$ of \emph{states}, 
	\item an initial state $\initstate \in \Config$,
	\item a finite set of variables $\PVar = \calo \cup \cals$,
	\item a \emph{labeling function} $V:\:\Config \rightarrow \distr(\cals)$,
	\item a \emph{transition relation} $\rightarrow \subseteq \Config \times \distr(\Config)$. 
\end{inparaenum}
\end{definition}
A PKS is \emph{fully probabilistic} if each state has at most one outgoing transition, i.e., if $c\rightarrow\mu$ and $c\rightarrow\mu'$ implies $\mu = \mu'$. We assume that programs always terminate.

\paragraph{Traces.}
A {\em trace} $T$ in $\calA$ is a sequence $T = c_0 c_1 c_2 \ldots c_n$ such that
\begin{inparaenum}[(i)]
	\item $c_i \in \Config, c_0 = \initstate$, and
	\item for all $0\leq i<n$, there exists a transition $c_i\rightarrow\mu$ with $\mu(c_{i+1})>0$. 
\end{inparaenum}
Let $\Trace(\calA)$ denote the set of traces of $\calA$.

\subsection{Probabilistic Schedulers}
A probabilistic scheduler is a function that implements a scheduling policy \cite{Sabelfeld99}, i.e., that decides with which probabilities the threads are selected. To be general, we allow a scheduler to use the full history of computation to make decision: given a path ending in some state $c$, a
scheduler chooses which of the enabled transitions in $c$ to execute \cite{Ngo13}.
Since each transition results in a distribution of states, a probabilistic scheduler returns a
distribution of distributions of states\footnote{Thus, we assume a discrete probability distribution over the uncountable set $\distr(\Config)$; only the countably many transitions occurring in $\calA$ can be scheduled with a positive probability.}.

\begin{definition}
A scheduler $\scheduler$ for $\calA=\langle \Config, \initstate, \PVar, \valuation, \rightarrow \rangle$ is a function
$\scheduler : \Trace(\calA) \rightarrow \distr(\distr(\Config))$,
such that, for all traces $T \in \Trace(\calA)$, $\scheduler(T)(\mu) > 0$ implies $\last(T) \rightarrow \mu$, where $\last(T)$ denotes the final state on $T$. 
\end{definition} 

The effect of a scheduler $\scheduler$ on a PKS $\calA$ can be described by a PKS $\calA_{\scheduler}$, i.e.,
the unreachable states of $\calA$ under the scheduler $\scheduler$ are removed by the transition relation $\rightarrow_\scheduler$.
Since all nondeterministic choices in $\calA$ have been resolved by $\scheduler$, $\calA_\scheduler$ is fully probabilistic. 
The probability $p(T)$ given to a trace $T = c_0 c_1 \ldots c_n$
is determined by $\delta(c_0)(c_1) \cdot \delta(c_0 c_1)(c_2)\cdot \cdot \cdot\delta(c_0 c_1 \ldots c_{n-1} )(c_n)$. 

To summarize, our approach assumes that the prior distribution of private data is uniform. This distribution represents the attacker's initial uncertainty about the private data. The program is considered a distribution transformer, from the initial distribution at the input to the final ones at the outputs. 

\subsection{Leakage of a Program Trace}\label{traceleak}
Example~\ref{interstate2} shows that the leakage of a program trace is not simply the sum of leakages of transition steps along the trace. This section addresses how we can compute the leakage of a program trace.
Consider again Example~\ref{interstate2},
\[
\begin{array}{rl}
& \calo:=0;  \\
& \calo:=\cals \:\&\: (001)_b;\\
& \calo:=\cals \:\&\: (011)_b;
\end{array}
\]
Let $(s_3 s_2 s_1)_b$ denote the binary form of $\cals$. The execution of this program results in only one trace, i.e., 
$\confi{(000)_b} \longrightarrow \confi{(00 s_1)_b} \longrightarrow \confi{(0 s_2 s_1)_b}$,
where a state $\confi{}$ is represented by the value of $\calo$ in that state.

Assume that $s_2=1$ and $s_1=1$, the obtained trace is
$\confi{(000)_b} \longrightarrow \confi{(001)_b} \longrightarrow \confi{(011)_b}$.

At the initial state $\confi{(000)_b}$, following our framework setting, i.e., the attacker only knows the possible values of private data, but he does not know which value is likely to be the true one, the attacker's initial uncertainty is represented by the uniform distribution of $\cals$, i.e., $\{0 \mapsto \frac{1}{8}, 1 \mapsto \frac{1}{8}, 2 \mapsto \frac{1}{8}, 3 \mapsto \frac{1}{8}, 4 \mapsto \frac{1}{8}, 5 \mapsto \frac{1}{8}, 6 \mapsto \frac{1}{8}, 7 \mapsto \frac{1}{8} \}$.
At the state $\confi{(001)_b}$, the attacker learns that the last bit of $\cals$ is $1$. Thus, the distribution of $\cals$ changes, for example, $\cals$ cannot be $0$. Hence, the updated distribution at state $\confi{(001)_b}$ is $\{0 \mapsto 0, 1 \mapsto \frac{1}{4}, 2 \mapsto 0, 3 \mapsto \frac{1}{4}, 4 \mapsto 0, 5 \mapsto \frac{1}{4}, 6 \mapsto 0, 7 \mapsto \frac{1}{4} \}$.
Similarly, at  final state $\confi{(011)_b}$, the updated distribution is $\{3 \mapsto \frac{1}{2}, 7 \mapsto \frac{1}{2} \}$\footnote{We leave out the elements that have probability $0$.}.

Based on the final distribution of $\cals$, the attacker derives that the value of $\cals$ is either $3$ or $7$. His uncertainty on secret information is reduced by the knowledge gained from the observation of the trace.

Since the program is a distribution transformer, the distribution of private data at the initial state of a trace can present the \emph{initial} uncertainty of the attacker about the secret, and the distribution of private data at the final state can present his \emph{final} uncertainty, after the trace has been observed.  Thus, we can define the leakage of a program trace as, 
\begin{center}
Leakage of a program trace = Initial uncertainty  - Final uncertainty.
\end{center}
The question arising is that how we represent the notion of uncertainty by a quantitative term.

Given a distribution of private data, the \emph{best} strategy of the one-try threat model is to choose the value with the maximum probability. `Best' means that this strategy induces the smallest probability of guessing wrong private data. Let $\cals$ be private data with carrier $\mathbf{S}$, the value that affects the notion of uncertainty is $\max_{s \in \mathbf{S}} p(s)$. If $\max_{s \in \mathbf{S}} p(s)=1$, the uncertainty must be $0$, i.e., the attacker already knows the value of $\cals$. Thus, the notion of uncertainty is computed as the negation of logarithm of $\max_{s \in \mathbf{S}} p(s)$, i.e.,
uncertainty = $- \log \:\: \max_{s \in \mathbf{S}} p(s)$,
where the negation is used to ensure the nonnegativeness property.

This measure coincides with the notion of R\'{e}nyi's min-entropy. Thus, given a distribution of $\cals$, the uncertainty of the attacker about the secret in our approach is:
Uncertainty = $\calH_{\renyi}(\cals)$.

Therefore, leakage of a program trace $T$ is,
\[\calL(T) = \calH_{\renyi}(\icals) - \calH_{\renyi}(\fcals),\]
where $\calH_{\renyi}(\icals)$ is R\'{e}nyi's min-entropy of $\cals$ with the initial distribution, and $\calH_{\renyi}(\fcals)$ is R\'{e}nyi's min-entropy of $\cals$ with the final distribution, i.e., the distribution of the secret at the final state in $T$.

Following our measure, in Example~\ref{interstate2}, $\calL(T)=- \log \frac{1}{8} - (- \log \frac{1}{2})=2$. This value matches the intuitive understanding that Example~\ref{interstate2} leaks 2 bits of the private data.

Consider again Example~\ref{interstate1}. Assume that $s_1=s_2=s_3=1$,
the execution of this program results in only one trace, i.e., 
$\confi{(000)_b} \longrightarrow \confi{(100)_b} \longrightarrow \confi{(011)_b}$.

At the state $\confi{(100)_b}$, the attacker learns that the first bit of $\cals$ is $1$. Thus, the distribution of $\cals$ is $\{4 \mapsto \frac{1}{4}, 5 \mapsto \frac{1}{4}, 6 \mapsto \frac{1}{4}, 7 \mapsto \frac{1}{4} \}$. 
At final state $\confi{(011)_b}$, the distribution is updated to $\{7 \mapsto 1 \}$, which is different from the final distribution in Example~\ref{interstate2}. Hence, $\calL(T)= - \log \frac{1}{8} - (- \log 1)=3$. This result also matches the intuition that the attacker is able to derive the value of $\cals$ precisely by observing the execution trace of Example~\ref{interstate1}. 

Notice that in our approach, instead of the notion of \emph{remaining} uncertainty, we use the notion of \emph{final} uncertainty. Both notions of initial and final uncertainty are denoted by the \emph{same} notion of entropy, i.e., R\'{e}nyi's min-entropy.  
The notion of remaining uncertainty depends only on the outcomes of the trace, while final uncertainty in our approach is based on the distribution computed for the final state, which takes into account the public values in the intermediate states along the trace, and also the program commands that result in such observables. This idea makes the following expression of program's leakage different from the one proposed by Smith \cite{Smith09}.


\subsection{Leakage of a Multi-threaded Program}
The execution of a multi-threaded program $P$ under the control of a scheduler $\scheduler$ often results in a set of traces, i.e., $\Trace(P)$. Therefore, the leakage of $P$ is computed as the \emph{expected} value of the leakages of its traces, i.e.,
\[
\begin{array}{rl}
\calL(P) & =  \sum_{T \in \Trace(\calA_{\scheduler})} p(T)\cdot \calL(T)\\

& = \sum_{T \in \Trace(\calA_{\scheduler})} p(T)(\calH_{\renyi}(\icals) - \calH_{\renyi}(\fcals)). \phantom{\bigg(}
\end{array}
\]

Since $\calH_{\renyi}(\icals)$ is the same for any $T \in \Trace(\calA_{\scheduler})$, thus for notational convenience, we simply write it as $\calH_{\renyi}(\iscals)$. We rephrase the above expression as follows,

\[\calL(P)=\calH_{\renyi}(\iscals) - \sum_{T \in \Trace(\calA_{\scheduler})} p(T)\cdot \calH_{\renyi}(\fcals).\]


Since the initial distribution of $\cals$ is uniform, $\calH_{\renyi}(\iscals)$ has the maximum value. Thus, $\calH_{\renyi}(\iscals) \geq \calH_{\renyi}(\fcals)$; then
$\calL(P)  \geq \calH_{\renyi}(\iscals) - \sum_{T \in \Trace(\calA_{\scheduler})} p(T) \cdot \calH_{\renyi}(\iscals) = 0$. 

Hence, following our approach, the computed leakage of any program is always non-negative, i.e., $\calL(P)\geq 0$ for all $P$. When
$\calL(P)=0$, the program is totally secure.

\subsection{A Case Study}\label{ananlysisprocess}
\begin{figure}[t]
\begin{center}
\begin{tikzpicture}[->,>=stealth',shorten >=1pt,auto,node distance=0.5cm,
                    thin][scale=0.8]
  \tikzstyle{every state}=[circle,draw,text=black,font=\sffamily\footnotesize]
\begin{scope}[every node/.style={scale=0.8}]
  \node[initial,state, initial text=\small{start}] at (9,7) [label=above:\small{\textbf{0}}] (1)         {0};
  \node[state]         (2) at (4,4) [label=above:\small{$\:\:\:\:\:\:$\textbf{1}}] {1};
  \node[state]         (3) at (9,5) [label=above:\small{$\:\:\:\:\:\:$\textbf{2}}] {0};
  \node[state]         (4) at (13,5) [label=above:\small{\textbf{3}}] {1};
  \node[state]         (5) at (2,2)      [label=above:\small{$\:\:\:\:\:\:$\textbf{4}}] {0};
  \node[state]         (6) at (6,2)      [label=above:\small {$\:\:$\textbf{5}}] {1};
  \node[state]         (7) at (9,3)      [label=above:\small{$\:\:\:\:\:\:\:\:$\textbf{6}}] {1};
  \node[state]         (8) at (13,3)      [label=above:\small{$\:\:\:\:\:\:$\textbf{7}}] {1};
  \node[state]         (9) at (0,0)      [label=above:\small{$\:\:\:\:\:\:\:$\textbf{8}}] {0};
  \node[state]         (10) at (1,0)     [label=above:\small{$\:\:\:\:\:\:\:$\textbf{9}}] {1};
  \node[state]         (11) at (2,0)     [label=above:\small{$\:\:\:\:\:\:\:\:\:\:\:\:\:\:\:$\textbf{10}}] {2};
  \node[state]         (12) at (3,0)     [label=above:\small{$\:\:\:\:\:\:\:\:$\textbf{11}}] {3};
  \node[state]         (13) at (4,0)      [label=above:\small{$\:\:\:\:$\textbf{12}}] {0};
  \node[state]         (14) at (5,0)     [label=above:\small{$\:\:\:\:\:\:\:\:\:\:$\textbf{13}}] {1};
  \node[state]         (15) at (6,0)     [label=above:\small{$\:\:\:\:\:\:\:\:$\textbf{14}}] {2};
  \node[state]         (16) at (7,0)     [label=above:\small{$\:\:\:\:\:\:\:\:$\textbf{15}}] {3};
  \node[state]         (17) at (8,1)     [label=above:\small{$\:$\textbf{16}}] {0};
  \node[state]         (18) at (10,1)     [label=above:\small{$\:\:\:\:\:\:$\textbf{17}}] {2};
  \node[state]         (19) at (12,1)     [label=above:\small{$\:$\textbf{18}}] {1};
  \node[state]         (20) at (14,1)     [label=above:\small{$\:\:\:\:\:\:$\textbf{19}}] {3}; 
  \path (1) edge        node [above left] {$\frac{1}{2}$}        (2)
            edge        node [above left] {$\frac{1}{4}$}        (3)
            edge        node [above right] {$\frac{1}{4}$}        (4)
        (2) edge        node [above left] {$\frac{1}{2}$}        (5)
            edge        node [above right] {$\frac{1}{2}$}        (6)  
        (3) edge                (7)
        (4) edge                (8)
        (5) edge        node [above left] {$\!\!\!\!\!\!\!\!\!\!\!\!\frac{1}{4}$}        (9)
            edge        node [above] {$\!\!\!\frac{1}{4}$}        (10)
            edge        node [above right] {$\!\!\!\frac{1}{4}$}        (11)
            edge        node [above right] {$\!\!\frac{1}{4}$}        (12)
        (6) edge        node [above left] {$\!\!\!\!\!\!\frac{1}{4}$}        (13)
            edge        node [above] {$\!\!\!\frac{1}{4}$}        (14)
            edge        node [above right] {$\!\!\!\frac{1}{4}$}        (15)
            edge        node [above right] {$\!\!\frac{1}{4}$}        (16)
        (7) edge        node [above left] {$\frac{1}{2}$}        (17)
            edge        node [above right] {$\frac{1}{2}$}        (18)
        (8) edge        node [above left] {$\frac{1}{2}$}        (19)
            edge        node [above right] {$\frac{1}{2}$}        (20);
      \end{scope}
\end{tikzpicture}
\end{center}
\caption{Model of the Program Execution}
\label{RunningExample}
\end{figure}
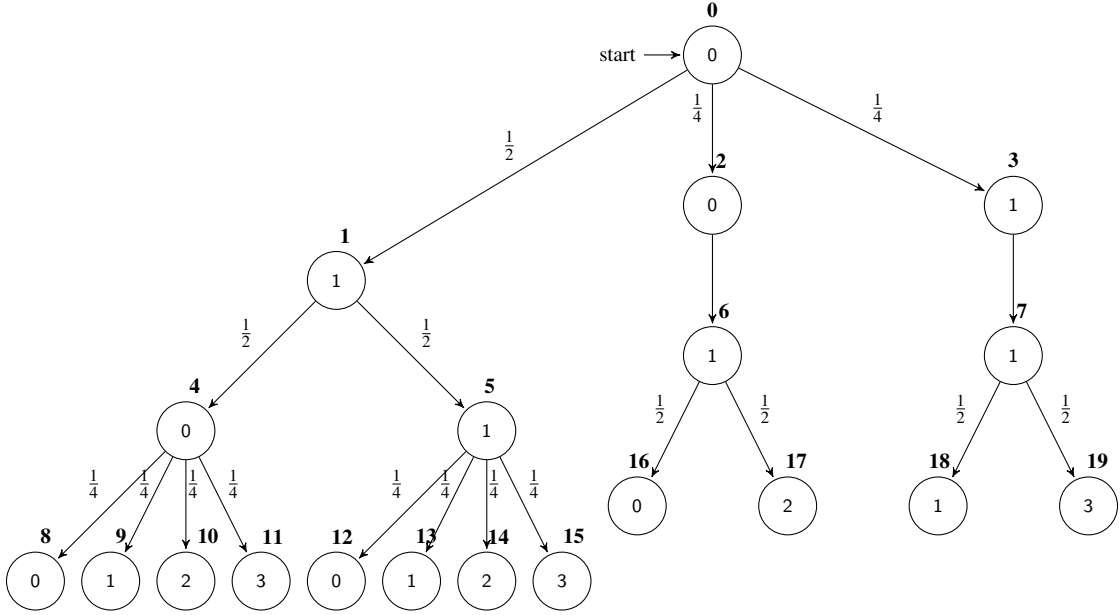

This part illustrates how the leakage of a multi-threaded program is computed. The following case study also shows that our measure is more precise than the measures given by the classical approaches. Consider the following example,
\begin{example}[Program $P_8$]
\[
\begin{array}{rl}
& \calo:=0;  \\
& \{\mathtt{if \: (\calo = 1)\:\:\:\:then\:\: \:\: \calo:=\cals/4\:\:\:\:else\:\:\:\:\calo:=\cals\!\!\!\!\! \mod 2\} \:\big|\!\big|\:\calo:=1;}\\
& \calo:=\cals\!\!\!\!\! \mod 4;
\end{array}
\]
where $\cals$ is a 3-bit unsigned integer. \label{casestudy} 
\end{example}
The execution of this program under the control of a uniform scheduler is illustrated by a PKS $\calA$ in Figure~\ref{RunningExample}. The PKS consists of $20$ states that are numbered from \textbf{0} (the initial state) to \textbf{19}. The content of each state is the value of  $\calo$ in that state, e.g., in the initial state, the value of $\calo$ is $0$, which corresponds with the first command of the program $\calo:=0$. 

Let $C_1$ and $C_2$ denote the left and right threads of the parallel composition operator. Since we consider the uniform scheduler, either thread $C_1$ or $C_2$ can be picked next with the same probability $\frac{1}{2}$. If the scheduler picks $C_2$ before $C_1$, $\calA$ evolves from state \textbf{0} to state \textbf{1}, where $\calo=1$. If $C_1$ is picked first, $\calA$ might evolve from state \textbf{0} to either state \textbf{2} or state \textbf{3} with the same probability $\frac{1}{4}$. Since the current value of $\calo$ is $0$, i.e., the value of $\calo$ in state \textbf{0}, the command $\calo:=\cals\!\!\!\!\! \mod 2$ is executed. Since the possible values of $\cals$ are $\{0,\ldots,7\}$, the outcome $\calo$ might be $0$ (state \textbf{2}) if $\cals \in \{0,2,4,6\}$, or $1$ (state \textbf{3}) if $\cals \in \{1,3,5,7\}$.

At state \textbf{1}, $\calA$ might evolve to either state \textbf{4} or state \textbf{5} with the same probability. Since currently, $\calo$ is $1$, the command $\calo:=\cals/4$ is executed. Thus, $\calo$ might be $0$ if $\cals \in \{0,1,2,3\}$, or $1$ if $\cals \in \{4,5,6,7\}$. 

The PKS $\calA$ evolves from one state to another until the execution terminates, i.e., when the last command $\calo:=\cals\!\!\!\!\! \mod 4$ is executed.

The initial uncertainty of the attacker about $\cals$ is denoted by the uniform distribution, i.e., $\{0 \mapsto \frac{1}{8}, 1 \mapsto \frac{1}{8}, 2 \mapsto \frac{1}{8}, 3 \mapsto \frac{1}{8}, 4 \mapsto \frac{1}{8}, 5 \mapsto \frac{1}{8}, 6 \mapsto \frac{1}{8}, 7 \mapsto \frac{1}{8} \}$. At state \textbf{1}, the distribution of $\cals$ is still uniform since the attacker learns nothing from the command $\calo:=1$. 
At state \textbf{4}, since the execution of $\calo:=\cals/4$ results in $0$, the attacker learns that the true value of $\calo$ must be in the set $\{0,1,2,3\}$. Thus, the updated distribution of $\cals$ at this state is $\{0 \mapsto \frac{1}{4}, 1 \mapsto \frac{1}{4}, 2 \mapsto \frac{1}{4}, 3 \mapsto \frac{1}{4}\}$. In the next step, the outcome of $\calo:=\cals\!\!\!\!\! \mod 4$ helps the attacker derive $\cals$ precisely, e.g., at state \textbf{8}, since $\calo=0$, the distribution of $\cals$ is $\{0 \mapsto 1\}$. Similarly, the attacker is also able to derive $\cals$ precisely, basing on the final distributions at states \textbf{9}, $\ldots$, \textbf{15}.  

At state \textbf{2}, the execution of $\calo:=\cals\!\!\!\!\! \mod 2$ results in $0$. Thus, the distribution of $\cals$ at this state is $\{0 \mapsto \frac{1}{4}, 2 \mapsto \frac{1}{4}, 4 \mapsto \frac{1}{4}, 6 \mapsto \frac{1}{4}\}$. This distribution remains unchanged at state \textbf{6}, since no information is gained from the execution of $\calo:=1$. At state \textbf{16}, the update distribution of $\cals$ is $\{0 \mapsto \frac{1}{2}, 4 \mapsto \frac{1}{2} \}$ since the execution of $\calo:=\cals\!\!\!\!\! \mod 4$ results in $0$. The same form of distributions is obtained at states \textbf{17}, \textbf{18}, \textbf{19}.

Among the $12$ possible traces, $8$ traces have the final uncertainty $0$, i.e., $\calH_{\renyi}(\fcals)= - \log 1 = 0$, and the other $4$ traces have the final uncertainty $1$, i.e., $\calH_{\renyi}(\fcals)= - \log \frac{1}{2} = 1$. The probability of traces with the final uncertainty $0$, i.e., traces end in state \textbf{8}, $\ldots$, \textbf{15}, is equal to the probability of traces with the final uncertainty $1$. Thus, according to our approach,
\[\calL(P_8) = 3 - (\frac{1}{2}\cdot 0 + \frac{1}{2}\cdot 1)= 2.5.\]

This value coincides with the real leakage of the program.
As we see, the last command $\calo:=\cals\!\!\!\!\! \mod 4$ always reveals the last $2$ bits of $\cals$. The first bit of $\cals$ might be leaked with the probability $\frac{1}{2}$, depending on whether the scheduler picks thread $C_2$ first or not. Thus, the real leakage of this program is $2.5$.

\subsection{Comparison}\label{multi-compare}
To reason about the exposed information flow of multi-threaded programs, Malacaria et al. \cite{Malacaria08} and Andr\'{e}s et al. \cite{Andres10} propose to reuse the classical information-theoretic approach but the observables are traces of public variables, instead of only the final outcomes. Consider the above case study. Since the possible values of $\cals$ are $\{0,\ldots,7\}$, the execution of $P_8$ results in the following traces,


\begin{center}
\begin{tabular}{|c|c|c|c|c|c|c|c|c|c|c|c|c|c|c|c|c|}
\hline
$\cals$             & \multicolumn{2}{|c|}{0} & \multicolumn{2}{|c|}{1} & \multicolumn{2}{|c|}{2} & \multicolumn{2}{|c|}{3} & \multicolumn{2}{|c|}{4} & \multicolumn{2}{|c|}{5} & \multicolumn{2}{|c|}{6} & \multicolumn{2}{|c|}{7} \\
\hline
\multirow{4}{*}{$\restrictto{T}{\calo}$} & 0 & 0 & 0 & 0 & 0 & 0 & 0 & 0 & 0 & 0 & 0 & 0 & 0 & 0 & 0 & 0 \\
& 0 & 1 & 1 & 1 & 0 & 1 & 1 & 1 & 0 & 1 & 1 & 1 & 0 & 1 & 1 & 1 \\
& 1 & 0 & 1 & 0 & 1 & 0 & 1 & 0 & 1 & 1 & 1 & 1 & 1 & 1 & 1 & 1 \\
& 0 & 0 & 1 & 1 & 2 & 2 & 3 & 3 & 0 & 0 & 1 & 1 & 2 & 2 & 3 & 3 \\
\hline
\end{tabular}
\end{center}
where $\restrictto{T}{\calo}$ denotes the trace of $\calo$.

By observing the traces of $\calo$, the attacker is able to derive information about $\cals$. For example,
if the obtained trace is $0100$, the attacker can derive $\cals$ precisely, since this trace is produced only when $\cals=0$. If the trace is $0010$, the attacker can conclude that $\cals$ is either $0$ or $4$ with the same probability, i.e., $\frac{1}{2}$. If the trace is $0111$, the possible value of $\cals$ is either $1$ or $5$, but with different probabilities, i.e., the probability of $\cals$ to be $5$ is $\frac{2}{3}$.

There are $6$ traces such that the attacker is able to derive the value of $\cals$ precisely from them. There are $4$ traces such that the attacker is able to guess $\cals$ correctly with the probability $\frac{1}{2}$, and $6$ traces with the probability $\frac{2}{3}$. Therefore,
\[\calL_{\Min}(P_8) = 3 - (- \log (\frac{6}{16} \cdot 1 + \frac{4}{16} \cdot \frac{1}{2} + \frac{6}{16} \cdot \frac{2}{3})) = 2.585.\] 
 
This computed leakage does not match the intuitive understanding, as we have just argued above. Notice that the classical approaches with the input-output model judge that this program leaks only $2$ bits of information, i.e., for each final outcome, the attacker might derive the secret with the probability $\frac{1}{2}$.

While the information-theoretic approach uses the notion of conditional min-entropy proposed by Smith \cite{Smith09} to denote the remaining uncertainty, we use the same notion of entropy to denote the initial and final uncertainty. The main difference is the position of the $\log$ in the expression of leakage. The idea of using logarithm is to express the notion of uncertainty in bits. Thus, the $\log$ should apply only to the distribution of private data which represents the uncertainty of the attacker, as in our approach. For multi-threaded programs, the distribution of traces depends strongly on schedulers, and thus we should distinguish between the distribution of traces and the distribution of private data. The notion of conditional min-entropy proposed by Smith~\cite{Smith09}, in which the logarithm applies to the combination of two distributions, do not distinguish between these two kinds of distribution, and thus it might cause imprecise results.

Notice that the expression $\sum_{T \in \Trace(\calA_{\scheduler})} p(T)\cdot \calH_{\renyi}(\fcals)$ is similar to the notion of conditional min-entropy defined by Cachin \cite{Cachin}. However, in our approach, we also take into account the effect of the scheduler, i.e., the transformation of distributions of private data depends also on which program commands producing the public values in the states. 
 
\subsection{Technique for Computing Leakage}
Another aspect is to develop a technique for computing the leakage of a multi-threaded program. The analysis consists of two steps. First, the execution of a multi-threaded program $P$ under the control of a given scheduler is modeled as a PKS $\calA$ in a standard way: The states of $\calA$ are tuples $\langle P,V \rangle$, consisting of a program fragment $P$ and a valuation $V:\:\Config \rightarrow \distr(\cals)$. The state transition relation $\rightarrow$ follows the small-step semantics of $C$. Based on the executed command and the obtained public result, the distribution of private data at each state is derived. We apply Kozen's  probabilistic semantics \cite{Kozen79} to present the transformation of probability distribution of $\cals$ during the program execution. 
The intuition behind a state transition is that it transforms an \emph{input} distribution of $\cals$ to an \emph{output} distribution so that the execution of a multi-threaded program results in a set of traces of probability distribution of $\cals$.

The computed leakage then follows trivially as the difference between the initial uncertainty and the expected value of the final uncertainties of all traces.  

\section{Related Work}\label{Relatedwork}
Several authors propose to use the concept from information theory to define leakage quantitatively. Most of the approaches are based on Shannon entropy and R\'{e}nyi's min-entropy \cite{Moskowitz03,Clark05,Chatzikokolakis06,Malacaria08,Malacaria10,Zhu05,Smith09,Andres10}. As illustrated above, these approaches might be counter-intuitive in some situations and also prone to conflicts when comparing between programs, i.e.,  Shannon-entropy measures judge this program more dangerous than the other one, but min-entropy measures give a counter-result. Thus, as argued by Alvim et al. \cite{Alvim12}, it seems that there is no unique measure that is likely to be appropriate in all cases.

To avoid the conflicts between classical approaches, Zhu et al. \cite{Zhu11} propose to view program execution as a probabilistic state transition from one state to another, where states denote probability distributions of the private data. This approach is close to our model in spirit, but their approach only aims to compare between programs. Their approach constructs probability distribution functions over the residual uncertainty about private data and then the comparison between programs is done by the means and the variances of the distributions.  


The quantitative security analysis proposed by Chen et al. \cite{Chen07} for multi-threaded programs defines the amount of leakage for each interleaving of the scheduler. The leakage of a program is then the expected value over all interleavings. The basis idea of this approach is that, for each interleaving of the scheduler, a set of possible traces is obtained. Based on the observables of the final states on these traces, the leakage of this interleaving is determined. This approach is imprecise since it does not consider leakages in intermediate states and the distribution of traces in a interleaving. 

Chen et al. \cite{Chen09} also define 
the leakage of a program trace as the sum of the products of the leakage generated by each transition step and the probability of the transition. Examples in Section~\ref{traceleak} show that this idea is not precise. This approach is only suitable to estimate the coarse maximum and minimum leakages of a program, i.e., the maximal and minimal values of leakages of traces. However, these values are not very helpful to judge programs, or to compare between different programs, because the gap between the maximum and minimum values is often large.

In another attempt to define the quantitative leakage for multi-threaded programs, Malacaria et al. \cite{Malacaria08} and Andr\'{e}s et al. \cite{Andres10} propose to reuse the classical information-theoretic approach but the observables are traces of public variables, instead of only the final outcomes. These approaches have been discussed in Section~\ref{multi-compare}.
Mu et al. \cite{MuC09} also model the execution of a probabilistic program by a probabilistic state transition system where states represent distributions of private data. Their paper proposes an automatic analyzer for measuring leakage. However, the analysis is only for sequential programs. 

\section{Conclusions and Future Work}\label{conclusion}
We propose a novel approach for estimating the leakage of a multi-threaded program. The notion of the leakage of a program trace is defined. The leakage of a program is then given as the expected value of the leakages of traces. Our approach takes into account the observables in intermediate states, and also the effect of the scheduler. The reasonableness of our approach should be further studied, in both theoretical properties and experimental case studies. However, we believe that our approach gives a more accurate way to study the quantitative security of multi-threaded programs, i.e., it agrees with the intuition about what the leakage should be. Thus, we consider this work as an important contribution in the field of quantitative security analysis for multi-threaded programs.

In the next step, the implementation of the analysis is planned. Since programs are modeled as PKSs, we are also considering some methods to avoid the state space explosion problem.

As mentioned in Section~\ref{OurApp}, the expression of the expected final uncertainty is very similar to the notion of conditional min-entropy defined by Cachin \cite{Cachin}. Thus,
we believe that our approach would coincide with the classical metric if we consider a channel where the observations are execution traces including scheduling decisions at steps along traces, and apply the Cachin's version of conditional min-entropy to the resulting channel. However, a formal proof needs to be done.

We also plan to study whether our approach also obtains good results for sequential programs. In this standard input-output model, 
if the initial distribution is non-uniform, the computed leakage might become negative. However, we believe that this negativeness property is only in the input-output setting. For multi-threaded programs, the value of leakage is always positive even when the priori distribution is non-uniform.
Thus, it is also valuable to study this property more. 

\vspace*{.5em}

\noindent \textbf{\large{Acknowledgments}.} The authors would like to thank Catuscia Palamidessi and Kostas Chatzikokolakis for many fruitful discussions. Our work is supported by NWO as part of the SlaLoM project.

\nocite{*}
\bibliographystyle{eptcs}
\bibliography{generic}
\end{document}